\documentclass[journal]{IEEEtran}
\usepackage{graphicx,float,epsfig,subfigure,enumerate,cite,bm,amsmath,color,soul,amsfonts}
\newcommand{\minew}[1]{{\color{black}{#1}}}

\ifCLASSINFOpdf
\else
\fi

\begin{document}

\title{Blockchain-Assisted Intelligent Symbiotic Radio in Space-Air-Ground Integrated Networks}

\author{Runze~Cheng,~\IEEEmembership{Student Member,~IEEE,}
        Yao~Sun,~\IEEEmembership{Senior~Member,~IEEE,} 
        Lina~Mohjazi,~\IEEEmembership{Senior~Member,~IEEE,}
        Ying-Chang~Liang,~\IEEEmembership{Fellow,~IEEE,}
        and~Muhammad~Imran,~\IEEEmembership{Senior~Member,~IEEE}

\thanks{Runze Cheng, Yao Sun, Lina Mohjazi, and Muhammad Ali Imran are with the James Watt School of Engineering, University of Glasgow, Glasgow G12 8QQ, U.K.}
\thanks{Ying-Chang Liang is with the Center for Intelligent Networking and Communications (CINC), University of Electronic Science and Technology of China (UESTC), Chengdu 611731, China.}
}
\maketitle

\begin{abstract}
    
        In a space-air-ground integrated network (SAGIN), managing resources for the growing number of highly-dynamic and heterogeneous radios is a challenging task.
        Symbiotic communication (SC) is a novel paradigm, which leverages the analogy of the natural ecosystem in biology to create a radio ecosystem in wireless networks that achieves cooperative service exchange and resource sharing, i.e., service/resource trading, among numerous radios. 
        As a result, the potential of symbiotic communication can be exploited to enhance resource management in SAGIN. 
        \minew{Despite the fact that different radio resource bottlenecks can complement each other via symbiotic relationships, unreliable information sharing among heterogeneous radios and multi-dimensional resources managing under diverse service requests impose critical challenges on trusted trading and intelligent decision-making.
        In this article, we propose a secure and smart symbiotic SAGIN ($S^4$) framework by using blockchain for ensuring trusted trading among heterogeneous radios and machine learning (ML) for guiding complex service/resource trading.} 
        A case study demonstrates that our proposed $S^4$ framework provides better service with rational resource management when compared with existing schemes. 
        Finally, we discuss several potential research directions for future symbiotic SAGIN.
    
\end{abstract}

\IEEEpeerreviewmaketitle

\section{Introduction}
The space-air-ground integrated network (SAGIN) is a promising architecture that incorporates the benefits of space, air, and terrestrial infrastructures to achieve the goal of ubiquitous coverage for mobile communication networks \cite{liu2018space}. 
Before fully achieving the potential of SAGIN, one critical challenge is to design an efficient resource management scheme that accommodates the network features of large scale, heterogeneity, and high dynamics.  
Meanwhile, the diversified service requirements make resource management more demanding.
\begin{figure*}[t]
    \centering
    \includegraphics[width=0.95\textwidth]{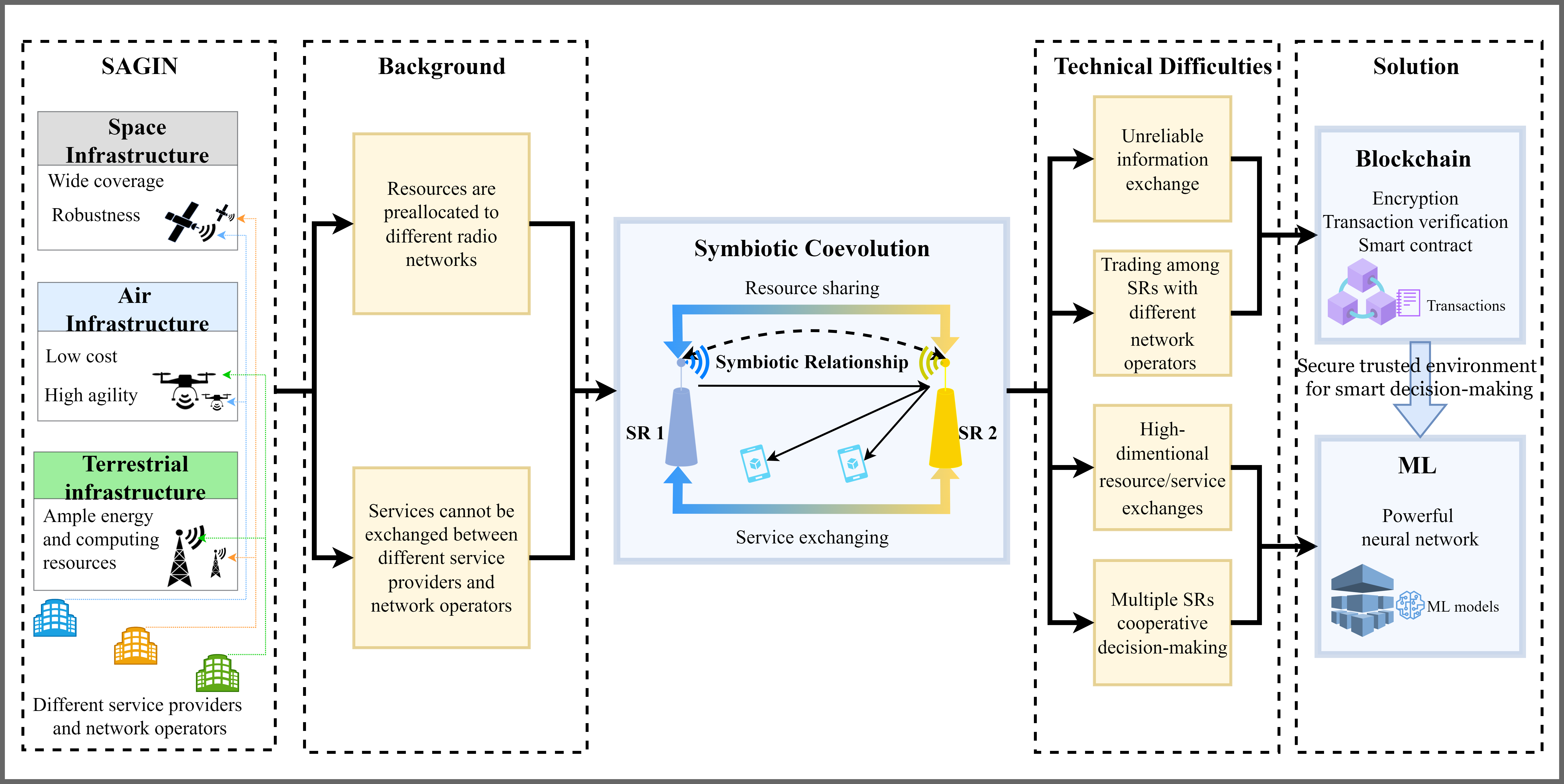}
    \caption{\minew{The challenges and motivations for symbiotic SAGIN.}} 
    \label{fig.s4}
\end{figure*}

Catering to the above challenges, symbiotic communication (SC) \cite{liang2022symbiotic} is introduced as a potential solution, which orchestrates different radios to share resources in a cooperative manner, thus achieving an enhanced resource utilization. 
\minew{Generally, SC borrows the analogy of biological symbiosis, where symbiotic relationships are formed among natural species, thus communication radios could establish symbiotic relationships via resource sharing and service exchange, i.e., service/resource trading \cite{liang2022symbiotic}.
Distinct from the traditional network where communication resources are pre-allocated and service exchanges among different service operators are not permitted, SC views the entire SAGIN as a radio ecosystem, where all symbiotic radios (SRs) achieve coevolution by service/resource trading \cite{janjua2021survey}. 
Therefore, different resource bottlenecks for individual SRs can complement each other by trading diversified resources and services. 
}

\minew{Nevertheless, two crucial technical challenges are highlighted in order to achieve the SC in SAGIN, i.e., trusted trading among various SRs and optimal decision-making of service/resource trading, as shown in Fig. \ref{fig.s4}.
In view of the long transmission distance, high network dynamics, and intricate electromagnetic interference, even in a hypothetical SAGIN that is not under attack, unreliable information exchanges may be extremely prevalent.
Additionally,  without an efficient consensus, it is challenging to achieve trusted trading because SRs in SAGIN could belong to various network operators.
Therefore, a trusted trading solution is required for an SC-enabled SAGIN.
Moreover, even when a trusted service/resource trading environment is secured, because heterogeneous SRs have various individual design objectives in terms of throughput, latency, dependability, etc., it is complicated to make mutually beneficial trading decisions for all SRs.
Meanwhile, a symbiotic relationship is typically built on a variety of resources and services, which adds to the complexity of the decision-making process.
}

\minew{
Fortunately, a promising trend that combines deep reinforcement learning (DRL) and blockchain can intuitively come to the rescue \cite{cheng2021blockchain}.
Blockchain can be employed as a digital ledger recording service exchanges in an immutable and traceable manner, thus establishing a trusted trading environment \cite{8668426}.
By encryption and transaction verification, only valid trading transactions can be recorded, which ensures the trust of coevolution among SRs \cite{hewa2020role}. 
Moreover, in order to achieve intelligent decision-making, machine learning (ML) can be exploited in the blockchain-secured trusted environment to efficiently guide SRs in optimal decision-making \cite{liu2018deep}.
With the help of powerful neural networks, ML can process the multidimensional and multivariate data produced by several SRs and determine the coevolutionary mechanism by which various services and resources are traded \cite{letaief2019roadmap}.
}

\minew{Accordingly, in this article, we propose a secure and smart symbiotic SAGIN ($S^4$) framework to achieve a highly efficient resource management mechanism, which is the first work to provide a visionary study in applying symbiotic communication to SAGIN.
In the $S^4$ framework, blockchain secures the precondition of coevolution, i.e., a trusted trading environment for heterogeneous SRs, while ML takes charge of intelligent decision-making for complicated service/resource trading during coevolution.
Especially, we discuss several technical challenges and the associated potential solutions to better exploit ML and blockchain in $S^4$ framework. 
Moreover, we demonstrate a case study to verify the superiorities of our proposed $S^4$ and discuss future research directions and challenges.
} 

The article is structured as follows. 
First, we present an overview of the $S^4$ network framework. 
Then, we discuss the two main technologies in $S^4$, namely blockchain and ML.
Furthermore, we present a case study where $S^4$ is implemented and provide numerical results accordingly. 
Finally, we discuss the future research directions of $S^4$ framework before conclusion.

\section{Symbiotic Communication in SAGIN}
In this section, we first briefly introduce SC with obligate and facultative symbiotic relationships. 
Subsequently, we elaborate on the proposed $S^4$ framework empowered by SC.

\begin{figure*}[t]
    \centering
    \includegraphics[width=0.9\textwidth]{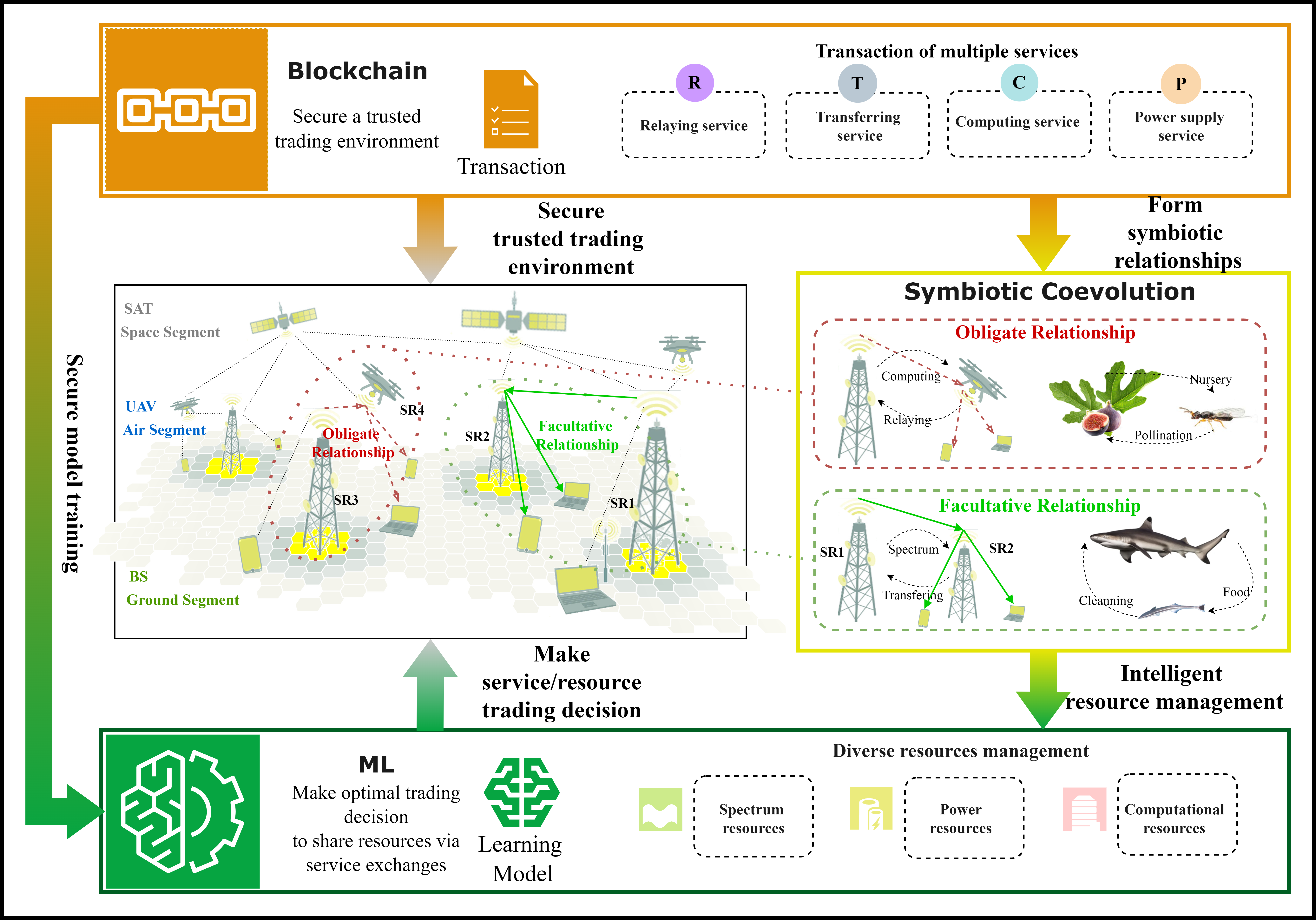}
    \caption{\minew{The system framework of $S^4$.}} 
    \label{fig.symbiotic}
\end{figure*}

\subsection{Preliminary: Symbiotic Radio Ecosystem}
Borrowing the idea from biology, each SR can be regarded as a specific biological species forming a radio ecosystem \cite{janjua2021survey}. 
While organisms consume resources (like food, light, and water), SRs as the biotic components of the ecosystem consume the radio resources (e.g., time, energy, and spectrum).
Moreover, analogous to species relationships (such as protection and phoresy) in the biological ecosystem, SRs can accomplish specific tasks and services (e.g., relaying, transmitting, computing) and interact with other SRs. 

Similary to the biological relationship, symbiosis is the long-term relationship among different SRs. 
Mutualism, amensalism, and parasitoidism are different forms of symbiotic relationships. 
In the SR ecosystem, mutualism is the relationship, in which both SRs benefit from each other. 
Meanwhile, amensalism is that when one SR receives benefits while the other is unaffected. 
Moreover, when one SR is harmed and the other receives benefits, it is called parasitoidism.
In symbiotic relationships, different connections can be built between SRs based on service/resource trading. 
In this work, we only discuss mutualism as a win-win relationship to efficiently manage resources.

Specifically, mutualism symbiosis can be obligate or facultative from a biological perspective. 
In SC, a facultative relationship is when multiple SRs cooperatively provide services for user equipments (UEs) to improve the service qualities \cite{liang2022symbiotic}. 
Meanwhile, it is acceptable for every SR to work as an independent server, which is analogous to the shark and remora relationship (Remoras collect more food like the scraps of prey dropped by the shark and parasites on the shark's skin and mouth, while the shark obtains cleaning service.)
For example, in Fig. \ref{fig.symbiotic}, both SR 1 (BS) and SR 2 (BS) are capable of independently providing network access services to UEs, while if one of them is unavailable, another BS can serve UEs alternatively.  
In an obligate relationship of SRs, one SR can provide services to UEs merely under the support of other SRs \cite{liang2022symbiotic}. 
Similar to the mutualism relationship between figs and fig-wasps (fig-wasps help pollinate figs while these insects cannot survive and reproduce without the living space, nursery, and nutrition provided by the figs), SR 3 (SAT) cannot transmit signals to terrestrial UEs without SR 4 (UAV) relaying signals as a satellite receiver, as shown in Fig. \ref{fig.symbiotic}.  

\subsection{$S^4$ Framework}
With respect to the features of symbiotic communications, we propose the $S^4$ framework for guiding SRs to trade services/resources cooperatively and establishing firm obligate or facultative relationships. 
\minew{In $S^4$, the objective is to improve resource utilization, as well as the service quality experienced by UEs via symbiotic coevolution.
The conception of symbiotic coevolution is cooperatively optimizing SRs' service/resource trading policies in evolutionary cycles to form stable obligate or facultative relationships.
In service/resource trading as shown in Fig. \ref{fig.symbiotic}, blockchain is used to establish a trusted trading environment via verifying transactions and evaluating the trustworthiness and activeness of SRs.
By deploying ML algorithms, SRs are capable of exchanging services according to the specific local environments, thus overcoming the resource bottlenecks of different clusters, as shown in Fig. \ref{fig.symbiotic}. 
With the help of trusted trading and intelligent decision-making, SRs can establish firm symbiotic relationships with appropriate neighbors. 
Accordingly, as resources are more efficiently managed across SRs via coevolution, the design objective can be accomplished. 
}

We consider multiple terrestrial UEs and various SRs belonging to the three segments of SAGIN, i.e., the satellite (SAT) segment, unnamed aerial vehicle (UAV) segment, and ground base station (BS) segment. 
As per the idea of symbiotic communications, all SRs can trade services/resources with neighbors and provide network services to UEs within their coverage.
In our $S^4$ framework, three categories of resources (i.e., spectrum, computing, and energy resources) are considered.
\minew{Supposing that two SRs (SR 1 and SR 2) cooperate with each other to share resources with the exchanges of the four services, thus forming different symbiotic relationships, as detailed below.
\begin{enumerate}[1)]
    \item Relaying service: When UEs cannot directly access service from the connected SR 1, SR 2 utilizes its spectrum resources to relay the radio signal that is sent from SR 1. This forms an obligate relationship.
    
    \item Transferring service: If SR 1 cannot provide the required network services to UEs, it transfers the network access task to SR 2, and the network access point changes from SR 1 to SR 2. Transferring service can lead to a facultative relationship.
    
    \item Computing service: The SR 2 supports computing tasks for SR 1, when SR 1 is struggling with inadequate computational or energy resources. Computing services can help to build either obligate or facultative relationships.
    
    \item Power supply service: SR 2 provides energy resources to ensure the general service of SR 1, which can be achieved via changing the power source, refueling the power source, simultaneous wireless power transfer, etc. Based on the power supply service, an obligate or facultative relationship can be formed between the two SRs. 
\end{enumerate} 
}

\section{The Role of Blockchain and ML in $S^4$} 
\begin{figure*}
    \centering
    \includegraphics[width=0.9\textwidth]{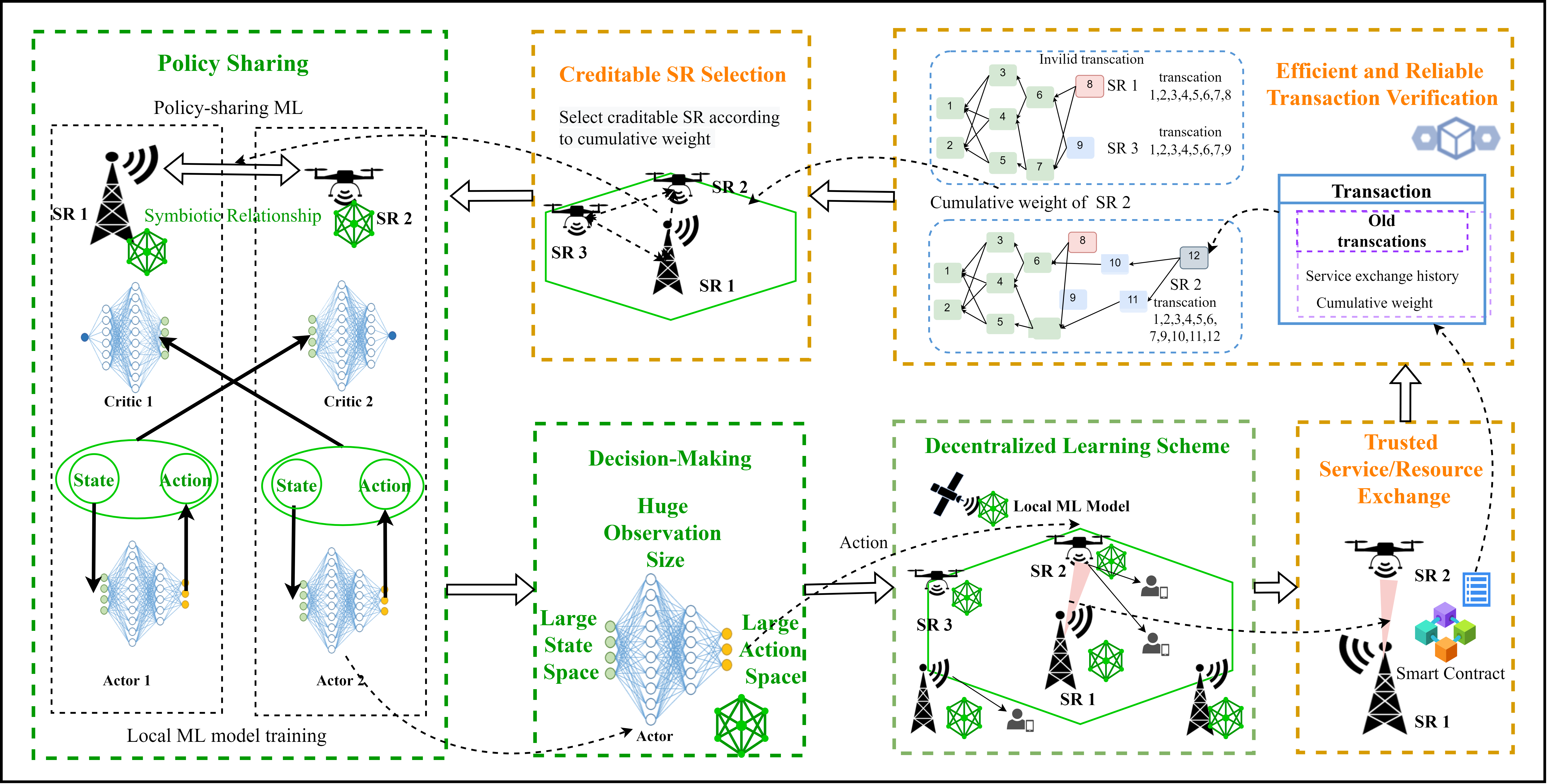}
    \caption{\minew{DAG-based blockchain consensus for policy-sharing ML.}}
    \label{fig.scheme}
\end{figure*}

In this section, we discuss in detail how the two technologies, namely, blockchain and ML can be exploited in the $S^4$ framework, as shown in Fig. \ref{fig.scheme}.

\subsection{Trusted Trading Environment Based on Blockchain}
In order to accomplish symbiotic coevolution, blockchain is exploited in the $S^4$ framework to secure a trusted trading and establish firm obligate or facultative relationships.
The main functions achieved by blockchain are as follows.

\textit{Trusted service/resource trading:} 
Since in SAGIN, SRs might belong to different network operators with different service providers, it is necessary to establish a trusted environment among SRs for service/resource trading. 
The smart contract, as a public agreement, directly sends the prices of services between SRs. 
Then, the balances of SRs' accounts are autonomously updated as per symbiotic service/resource trading without third-party involvement, which ensures all the service/resource transactions are traceable, transparent, and irreversible. 

\textit{Efficient and reliable transaction verification:} 
In $S^4$, pervasive connectivity is accomplished among SRs, which brings a ubiquitous symbiotic service/resource trading. 
In addition, the information sharing in $S^4$ tends to be time-sensitive.
Therefore, high scalability and fast transaction verification are required in the blockchain used in $S^4$.
\minew{To meet the requirements, as a kind of data structure that assembles the relations of transactions as tree logic, the directed acyclic graph (DAG)-based blockchain with high transaction throughput is a promising solution. 
By applying a gossip algorithm in the DAG-based blockchain, SRs have the right to record and broadcast their transactions after verifying and packaging a certain number of neighbors' transactions \cite{xie2021resource}.
Because complex hash calculations and global verifications are not essential for DAG-based blockchain, SRs can mutually secure each other and verify neighbors' transactions in a short time \cite{park2019performance}. 
Moreover, in the DAG-based blockchain, all newly created transactions must acknowledge several existing transactions via verification and reference \cite{cui2019efficient}, as shown in Fig. \ref{fig.scheme}.
For transactions that are built by malicious nodes, the probability of being ignored by other SRs is increasing.
As fake and fault transactions are not preserved, DAG-based blockchain can contribute to trusted trading.
}

\textit{Creditable SR selection:} A concept of cumulative weight can be used to measure the trustworthiness and activeness of different SRs, thus helping SRs find creditable policy-sharing partners.
The cumulative weight of an SR is defined as the sum of the trustworthiness weights of its all transactions, while the trustworthiness weights of a transaction are issued by SRs who agree with it and are proportional to the consumed computing and energy resources for verification. 
Therefore, if an SR maintains a higher cumulative weight, it means this SR has more verified transactions and its transactions are trusted by more SRs.
Through sharing policies with these creditable SRs with high cumulative weights, it is efficient to enhance the obligate and facultative symbiotic relationships and accomplish symbiotic coevolution. 

\subsection{Intelligent Resources Management in Coevolution}
ML should be utilized for guiding SRs to cooperatively make optimal service/resource trading decisions for intelligent resource management in symbiotic coevolution \cite{bithas2019survey}. 
However, applying ML to the $S^4$ framework introduces a number of challenges due to SAGIN's large-scale, heterogeneity, and high-dynamics characteristics, as well as complicated service/resource exchanges in SC \cite{kato2019optimizing}.

\textit{Centralized/decentralized learning scheme:} One fundamental issue to discuss is whether to use centralized or decentralized learning schemes to achieve optimal service/resource trading.
In the $S^4$ framework, centralized learning schemes require collecting all SRs' observed data to train a global resource management policy, while in decentralized schemes, each SR acts as an agent to train and maintain its own policy. 
Due to the large number of SRs in SAGIN, tackling complex states with a single centralized model is extremely difficult, resulting in a severe learning performance degradation.
Furthermore, some UAVs might require multi-hop transmissions to get decisions from the central controller, which may incur the concern on delay and reliability.
Therefore, in the $S^4$ framework, it is preferable to deploy decentralized learning schemes on SRs to decide service/resource trading, as demonstrated in Fig. \ref{fig.scheme}. 
Even partial SRs do not have enough computing and energy resources to support the intensive computation of neural networks, the resource limitations on these SRs can be effectively addressed via symbiotic service/resource trading.

\textit{Huge observation size:} To perform model training in the $S^4$ framework, SRs should observe their environments and determine the states. 
Normally, SRs need both its own and other SRs' observations to better understand their present state. 
To avoid the huge amount of shared data, only crucial information such as location and resources allocation should be shared among the nearby SRs.
Moreover, due to geographical isolation, interactions with SRs in different regions may be limited, even for the SRs from the same SAGIN segment.
Accordingly, rather than sharing data with all SRs in the SAGIN, SRs should emphasize observation sharing with their nearby neighbors.

\textit{Large state space:} With respect to the massive SRs and multiple types of resources in $S^4$, designing intelligent service/resource trading schemes normally requires a large state space to precisely describe the SAGIN network environment. 
Therefore, exploiting deep Q network (DQN)-based or actor-critic-based algorithms is attractive. 

\textit{Large action space:} In the $S^4$ framework, the action of an SR should be service/resource trading decisions that might include the type of shared resources, the amount for sharing, and the target SR shared with.
Therefore, in this symbiotic SAGIN scenario, high-dimensional action spaces should be necessary within policy learning.
To address this challenge, modeling SR policy using a multivariate distribution and developing multiple separate learning models are identified as two promising methods.
However, for granular resource allocation, continuous action space might still result in ultra-large action space issues.
Accordingly, DQN-based algorithms, which are often employed to solve discrete action space problems, could be ineffective in $S^4$, whereas actor-critic is a viable alternative.

\textit{Reward design:} In an intelligent coevolution scheme, rewards can be obtained by providing any kind of the four services discussed in Section II.
In local model learning, to avoid SRs only focusing on their own rewards, two methods are generally utilized.
One is to use the entire system's global reward in local model training, which indicates that all SRs use the same average reward; another method is to design the same reward calculation function for all SRs that produces distinct rewards for each SR based on their actions and corresponding environment.
Because an SR's environment includes only itself and its neighbors, rather than all SRs, the second method should be more appropriate in the $S^4$ framework.

\textit{Policy contradiction:} Contradiction can exist between the policies of different SRs due to the partial observation.
Therefore, a policy-sharing ML is needed in the $S^4$. 
According to the facultative and obligate symbiotic relationships, SRs are capable of determining which neighbors' policies might influence them most. 
Afterwards, SRs can save and use the newest policies shared by their related creditable neighbors in local model training, thus avoiding conflicts of interest.

\section{Implementations of $S^4$}
\begin{figure*}[t]
    \centering
    \includegraphics[width=0.9\textwidth]{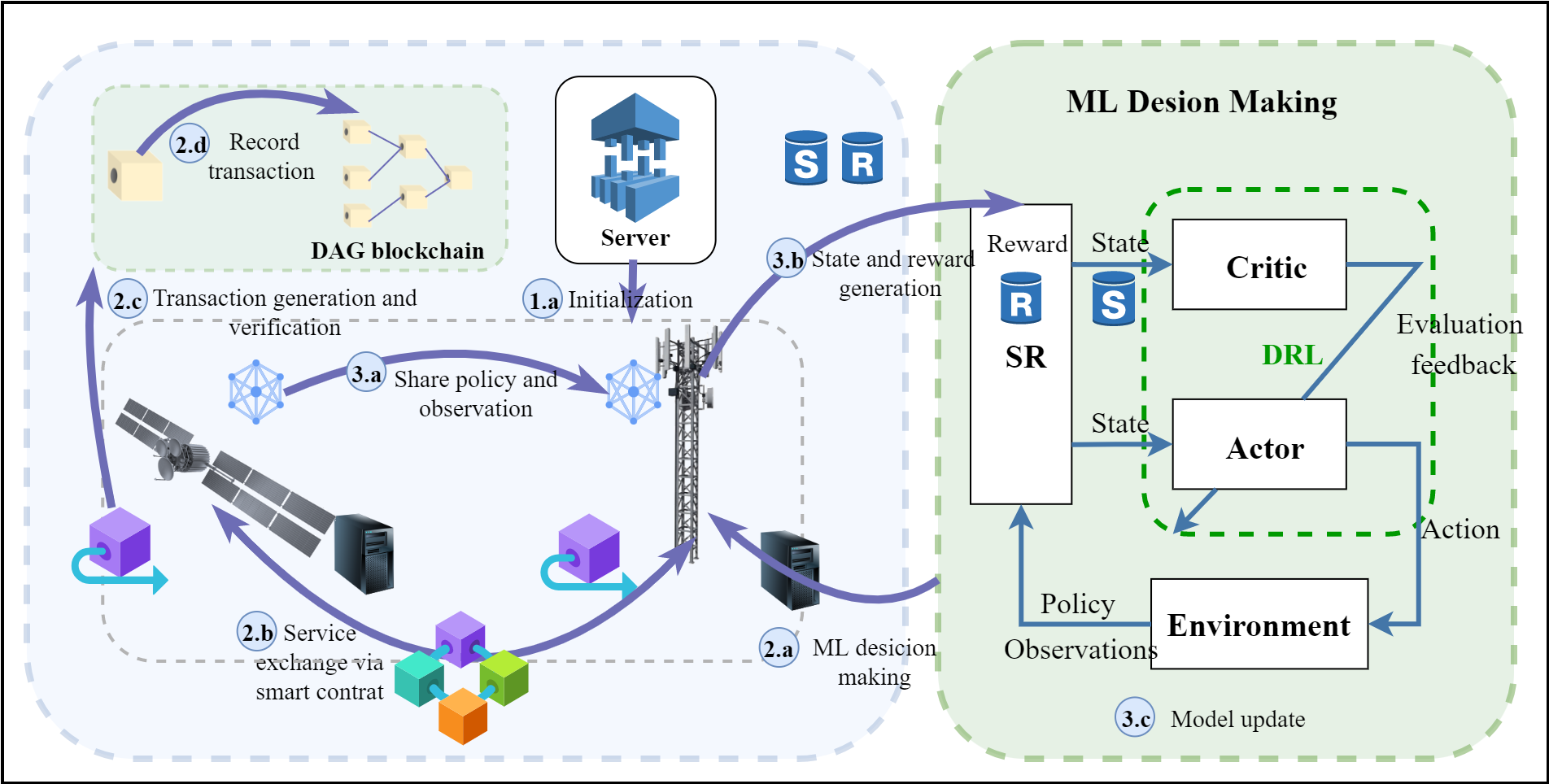}
    \caption{Simple flow chat of DAG blockchain consensus and policy-sharing ML implementation.}
    \label{fig.flow}
\end{figure*}

\minew{In this section, we present how blockchain and policy-sharing ML are operated in the $S^4$ framework.
We assume all the SRs are with adopted resources to participate in executing the blockchain consensus mechanism, cooperative policy learning and distributed decision-making.} 
Basically, it contains three steps, which are 1) initialization, 2) service/resource trading and transaction verification, and 3) policy sharing and model update, as shown in Fig. \ref{fig.flow}.

\subsection{Step 1: Initialization}
Initially, the SAT network starts the SC participation verification. 
Then, SRs send their information, like location, radio type, power supply type, network condition, etc. via the SAT communication link to apply for participation. 
Subsequently, the SAT controller decides which agents meet the minimum requirements of service/resource trading, and sends the initialization information to these SRs. 
The initialization information includes the architecture of the policy-sharing ML model, DAG blockchain transaction generate principle, reward calculation function, all SRs' and UEs' locations, etc. 
Although the same information is shared to every SR, the observations and learning models generated by different SRs are not the same, leading to different policies. 

\subsection{Step 2: Service Exchanging and Transaction Recording} 
In the decision-making step, SRs allocate communication resources based on actions (step 2.a in Fig. \ref{fig.flow}). 
Subsequently, the DAG consensus execution process starts, where SRs begin to trade services/resources via smart contract and cooperatively provide network access services for UEs according to the received network access requests in their coverage area (step 2.b).
The unverified dataset like service/resource trading records, the state of the environment, and cumulative trustworthiness weights will be first preserved in transactions locally by SRs.
By using a gossip algorithm, SRs can broadcast their local transaction to all neighbor SRs.
Once the neighbors receive a transaction, they will check the signature and verify the validity of this transaction as verifier SRs (step 2.c).
The trustworthiness weight of this transaction will be issued by verifier SRs in proportion to the consumed computing and energy resource.
After that, this transaction can be packed and recorded into the head of a new transaction generated by verifier SRs. 
In the end, only the longest chain should be reserved while others are deleted as per the longest chain rule \cite{park2019performance} (step 2.d). 

\subsection{Step 3: Policy Sharing and Model Updating}
First, the cumulative trustworthiness weight of an SR can be calculated based on its transactions' trustworthiness weights. 
Then, SRs share their own latest policies with neighbors and save the policies from neighbor SRs with high cumulative weights as reference. 
After that, SRs get partial corresponding observations from neighbors (step 3.a), and integrate their individual observations and their neighbors' observations to determine their own states. 
Meanwhile,  SRs use the reward calculation function to access their action reward based on validated service/resource trading data (step 3.b). 
Finally, as learning information like action, state, reward, and relevant neighbor policy are fetched, SRs begin to train and update their local models (step 3.c). 

\section{Case Study}
In this section, simulations of a case study are presented to evaluate the performance of the $S^4$ based resource management scheme. 
\minew{Based on the aforementioned rules in Section III for choosing ML algorithm and blockchain consensus, the resource management scheme in $S^4$ of this case study is DAG-based blockchain empowered multi-agent proximal policy optimization (MAPPO) \cite{yu2021surprising}.
A DAG consensus algorithm inspired from Tangle \cite{popov2018tangle}, is used in our blockchain to ensure a high transaction throughput with a low resource consumption since it can process amounts of transactions simultaneously in an asynchronous manner without mining and complicated hash calculation.
Meanwhile, a modified MAPPO promotes symbiotic coevolution because it allows SRs to learn the approximate policies of their creditable symbiotic cooperators, which avoids conflicts in policies. 
For comparisons, the three benchmarks are as follows.

\begin{enumerate}
    \item Non-SC: No symbiotic service/resource trading, resources are pre-allocated, rest parts are the same as $S^4$ (Benchmark for verifying the effectiveness of SR).
    \item Non-blockchain: Similar to $S^4$, except DAG-based blockchain is disabled (Benchmark for verifying the effectiveness of blockchain).
    \item Non-ML: The policy is giving priority to SRs with high cumulative weights, rest parts are the same as $S^4$ (Benchmark for verifying the effectiveness of ML).
\end{enumerate}
}

\subsection{Simulation Settings}
In our simulation, we consider a SAGIN containing 45 BSs, 4 UAV clusters (each UAV cluster contain 10 UAVs), and 1 SAT. 
There are 200 UEs with two hot spots in the considered area.
Meanwhile, the UAV clusters are uniformly distributed within the area, while the SAT is $1000km$ above the ground.
Moreover, the radius of the coverage area for a BS and a UAV cluster are set as $200m$ and $100+150m$ (transmission distance and flight radius), respectively. 
In this study, only spectrum resource is considered, where the bandwidth of a BS, a UAV, and a SAT are set to $20MHz$, $10MHz$, and $500MHz$, respectively.
We assume all UEs are fixed, but with a dynamic service requirement, where the transmission rate requirement range is $ \left[10,1000\right] Mbps$, and the range of latency is $\left[1,100\right] ms$. 

In the proximal policy optimization (PPO)-based local model, an actor network has an input layer with 8 neurons and a single neural output layer, as well as 2 hidden layers with 256 neurons.
We employ $ReLU$ as the activation function between the input layer and the two hidden layers, and $Softmax$ as the activation function between the second hidden layer and the output layer.
The critic network is the same as the actor network except that there is no activation function between the second hidden layer and the output layer.
The learning rate $\alpha $ is set as 0.0003, and mini-batches of one training iteration are set to 16.
\subsection{Numerical Results}
\begin{figure}[t]
    \centering
    \includegraphics[width=0.45\textwidth]{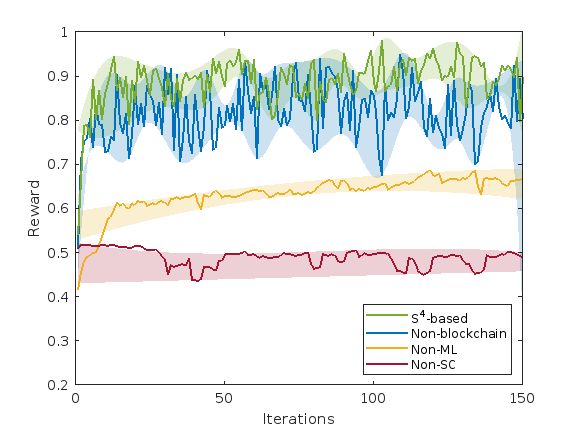}
    \caption{Convergence performance for different schemes}
    \label{fig.Sim1}
\end{figure}

We first show the reward convergence  for the four schemes, as depicted in Fig. \ref{fig.Sim1}.
The reward is proportional to UEs' satisfaction with the maximum value 1.
It is noted that fault/fake transactions can be randomly generated by an SR and might make chaos in training. 
Obviously, the proposed $S^4$-based scheme achieves the highest reward of around 0.9 in 25 training iterations. 
Without securing trusted trading by the DAG-based blockchain, the Non-blockchain scheme attains the reward of around 0.85 in 50 iterations with a large variation in rewards. 
The Non-ML scheme achieves a reward of about 0.6 because the distributed heuristic algorithm cannot guide SRs in effective cooperation, whereas the rewards of the Non-SC scheme are all below 0.55 since there is no symbiotic cooperation.

\begin{figure}[t]
    \centering
    \includegraphics[width=0.45\textwidth]{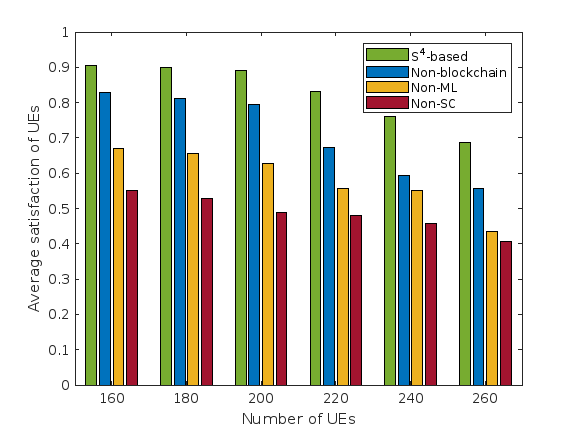}
    \caption{UE's satisfaction ratio for different schemes.}
    \label{fig.Sim2}
\end{figure}

Fig. \ref{fig.Sim2} compares the average satisfaction ratios under different number of UEs for four schemes. 
It is noted that UE's satisfaction is the weighted average ratio of each service quality value required by UE and its corresponding actual experienced quality value, where the service quality factors including delay and transmission rate.
We observe that the proposed $S^4$-based scheme always outperforms the other three schemes, and it keeps a satisfaction ratio of about 0.9 when the number of UEs is lower than 200.
Because of intelligent decision-making, the proposed $S^4$-based scheme can guarantee the service requirements of more UEs compared with the Non-ML and Non-SC schemes.
Moreover, without the trusted trading secured by blockchain, the Non-blockchain scheme attains a lower satisfaction ratio than that of $S^4$.

\minew{
\section{Challenges and Future Directions}
This article introduces a fundamental framework $S^4$ for intelligent resource management for the SAGIN. 
To unlock the full potential of the $S^4$, some challenges and research directions are discussed for further exploration in the future. 

\subsection{Diversified Symbiotic Relationships}
In the current $S^4$ framework, we only investigate the mutualism relationship, which may not accurately describe the wireless communication system in every application scenario. 
Other symbiotic relationships might be co-existed because a symbiotic system generally cannot only benefit from cooperation but also from competition.
To form different symbiotic relationships, SRs' objectives might need to alter between self-interest and altruism according to environments.
Therefore, the adjustable symbiotic coevolution objective requires further research to exploit diversified symbiotic relationships. 

\subsection{Signaling Storm in SC}
In $S^{4}$, signaling overhead is occurred in ML model training and the service/resources trading process, especially the trading process may incur massive signaling. 
SRs should cooperate to provide service for UEs, which might require a huge volume of signaling exchanges not only between massive UEs and numerous SRs but also among SRs. 
In this case, signaling storm may happen when the service exchanges are extremely frequent in an SR ecosystem, which can overwhelm network resources and lead to intolerant service outages. 
Adjusting the coevolutionary cycle and resource trading period based on the local model convergence and the formed symbiotic relationships could relieve the burden on signaling overhead in SR ecosystems.

\subsection{Functionality Modularity of SRs}
In $S^4$, we assume all the SRs are capable of participating in both ML and blockchain systems, which may lead to an over-complexity of the system and unnecessary waste of computing and energy resources. 
For example, the SR without sufficient computing power needs to pay extra cost to get computing support from other SRs via service/resources trading to participate in blockchain/ML.
To avoid this, SR functionality modularity is a promising remedy, in which SRs are divided into different functionality groups based on computing capability, energy resources, storage size, etc. so that different SRs can take charge of appropriate tasks. 

\subsection{Fast Convergence in Distributed Learning}
For the practical deployment of distributed cooperative learning, accelerating the convergence of model training is challenging. 
The learning models in our $S^4$ are updated asynchronously because of the difference in computing resources among SRs. 
However, since each individual policies in $S^4$ affect each other during policy-sharing ML model training, the models converged slowly under low computing resources may delay the convergence of other SRs' models. 
Moreover, in $S^4$, all newly participated SRs have to train their models from scratch, which also causes the low convergence speed. 
To solve that, federated learning, transfer learning and meta learning might be promising directions to speed the convergence by sharing model parameters.}

\section{Conclusion}
In this article, we proposed an $S^4$ framework that introduces symbiotic communication into SAGIN and that is powered by ML and blockchain to achieve intelligent resource management via secure service trading. 
Simulation results of a case study demonstrated the superiorities of the $S^4$ framework, in terms of communication resources utilization and the quality of service provided to UEs.
In addition, a forward-looking vision on future research directions is offered.
In general, we expect this work to be a pioneer in setting the steppingstone for intelligent resource management scheme for a symbiotic SAGIN with the interplay of ML and blockchain.

\bibliographystyle{IEEEtran}
\bibliography{IEEEabrv,bibSCMagazine}

\begin{thebibliography}{10}
\providecommand{\url}[1]{#1}
\csname url@samestyle\endcsname
\providecommand{\newblock}{\relax}
\providecommand{\bibinfo}[2]{#2}
\providecommand{\BIBentrySTDinterwordspacing}{\spaceskip=0pt\relax}
\providecommand{\BIBentryALTinterwordstretchfactor}{4}
\providecommand{\BIBentryALTinterwordspacing}{\spaceskip=\fontdimen2\font plus
\BIBentryALTinterwordstretchfactor\fontdimen3\font minus
  \fontdimen4\font\relax}
\providecommand{\BIBforeignlanguage}[2]{{%
\expandafter\ifx\csname l@#1\endcsname\relax
\typeout{** WARNING: IEEEtran.bst: No hyphenation pattern has been}%
\typeout{** loaded for the language `#1'. Using the pattern for}%
\typeout{** the default language instead.}%
\else
\language=\csname l@#1\endcsname
\fi
#2}}
\providecommand{\BIBdecl}{\relax}
\BIBdecl

\bibitem{liu2018space}
J.~Liu, Y.~Shi, Z.~M. Fadlullah, and N.~Kato, ``{Space-Air-Ground Integrated
  Network: A Survey},'' \emph{IEEE Communications Surveys \& Tutorials},
  vol.~20, no.~4, pp. 2714--2741, 2018.

\bibitem{liang2022symbiotic}
Y.-C. Liang, R.~Long, Q.~Zhang, and D.~Niyato, ``{Symbiotic Communications:
  Where Marconi Meets Darwin},'' \emph{IEEE Wireless Communications}, vol.~29,
  no.~1, pp. 144--150, 2022.

\bibitem{janjua2021survey}
M.~B. Janjua and H.~Arslan, ``{Survey on Symbiotic Radio: A Paradigm Shift in
  Spectrum Sharing and Coexistence},'' \emph{arXiv preprint arXiv:2111.08948},
  2021.

\bibitem{cheng2021blockchain}
R.~Cheng, Y.~Sun, Y.~Liu, L.~Xia, D.~Feng, and M.~A. Imran,
  ``{Blockchain-Empowered Federated Learning Approach for an Intelligent and
  Reliable D2D Caching Scheme},'' \emph{IEEE Internet of Things Journal},
  vol.~9, no.~11, pp. 7879--7890, 2021.

\bibitem{8668426}
Y.~Sun, L.~Zhang, G.~Feng, B.~Yang, B.~Cao, and M.~A. Imran,
  ``{Blockchain-Enabled Wireless Internet of Things: Performance Analysis and
  Optimal Communication Node Deployment},'' \emph{IEEE Internet of Things
  Journal}, vol.~6, no.~3, pp. 5791--5802, 2019.

\bibitem{hewa2020role}
T.~Hewa, G.~G{\"u}r, A.~Kalla, M.~Ylianttila, A.~Bracken, and M.~Liyanage,
  ``{The Role of Blockchain in 6G: Challenges, Opportunities and Research
  Directions},'' in \emph{2020 2nd 6G Wireless Summit (6G SUMMIT)}.\hskip 1em
  plus 0.5em minus 0.4em\relax IEEE, 2020, pp. 1--5.

\bibitem{liu2018deep}
M.~Liu, T.~Song, and G.~Gui, ``{Deep Cognitive Perspective: Resource Allocation
  for NOMA-Based Heterogeneous IoT with Imperfect SIC},'' \emph{IEEE Internet
  of Things Journal}, vol.~6, no.~2, pp. 2885--2894, 2018.

\bibitem{letaief2019roadmap}
K.~B. Letaief, W.~Chen, Y.~Shi, J.~Zhang, and Y.-J.~A. Zhang, ``{The Roadmap to
  6G: AI Empowered Wireless Networks},'' \emph{IEEE Communications Magazine},
  vol.~57, no.~8, pp. 84--90, 2019.

\bibitem{xie2021resource}
J.~Xie, K.~Zhang, Y.~Lu, and Y.~Zhang, ``{Resource-Efficient DAG Blockchain
  with Sharding for 6G Networks},'' \emph{IEEE Network}, 2021.

\bibitem{park2019performance}
S.~Park, S.~Oh, and H.~Kim, ``{Performance Analysis of DAG-Based
  Cryptocurrency},'' in \emph{2019 IEEE International Conference on
  Communications workshops (ICC workshops)}.\hskip 1em plus 0.5em minus
  0.4em\relax IEEE, 2019, pp. 1--6.

\bibitem{cui2019efficient}
L.~Cui, S.~Yang, Z.~Chen, Y.~Pan, M.~Xu, and K.~Xu, ``{An Efficient and
  Compacted DAG-based Blockchain Protocol for Industrial Internet of Things},''
  \emph{IEEE Transactions on Industrial Informatics}, vol.~16, no.~6, pp.
  4134--4145, 2019.

\bibitem{bithas2019survey}
P.~S. Bithas, E.~T. Michailidis, N.~Nomikos, D.~Vouyioukas, and A.~G. Kanatas,
  ``{A Survey on Machine-Learning Techniques for UAV-Based Communications},''
  \emph{Sensors}, vol.~19, no.~23, p. 5170, 2019.

\bibitem{kato2019optimizing}
N.~Kato, Z.~M. Fadlullah, F.~Tang, B.~Mao, S.~Tani, A.~Okamura, and J.~Liu,
  ``{Optimizing Space-Air-Ground Integrated Networks by Artificial
  Intelligence},'' \emph{IEEE Wireless Communications}, vol.~26, no.~4, pp.
  140--147, 2019.

\bibitem{yu2021surprising}
C.~Yu, A.~Velu, E.~Vinitsky, Y.~Wang, A.~Bayen, and Y.~Wu, ``{The Surprising
  Effectiveness of PPO in Cooperative, Multi-Agent Games},'' \emph{arXiv
  preprint arXiv:2103.01955}, 2021.

\bibitem{popov2018tangle}
S.~Popov, ``{The Tangle},'' \emph{White Paper}, vol.~1, no.~3, 2018.

\end{thebibliography}

\end{document}